\documentclass[useAMS,usenatbib]{mn2e}
\usepackage{epsfig}
\usepackage{rotating}
\usepackage{lscape,graphicx}
\usepackage{color}
\usepackage{natbib}
\usepackage{amsmath}
\usepackage{comment}

\usepackage{color}

\newcommand{\cmjj}{\mbox{${\rm cm^{-2}}$}}

\newcommand{\msun}{\mbox{${\rm M}_\odot$}}
\providecommand{\kms}{\,\ensuremath{\rm{km\,s}^{-1}}}

\newcommand\mnras{{MNRAS}}%
\newcommand\araa{{ARA\&A}}%
\newcommand\apj{{ApJ}}%
\newcommand\apjl{{ApJ}}%
\newcommand\aap{{A\&A}}%
\newcommand\aapr{{A\&A~Rev.}}%
\newcommand\aaps{{A\&AS}}%
\newcommand\aj{{AJ}} %
\newcommand\pasp{{PASP}} %
	 
\def\lesssim{\mathrel{\hbox{\rlap{\hbox{\lower3pt\hbox{$\sim$}}}\hbox{\raise2pt\hbox{$<$}}}}}
\def\gtrsim{\mathrel{\hbox{\rlap{\hbox{\lower3pt\hbox{$\sim$}}}\hbox{\raise2pt\hbox{$>$}}}}}	 

\title[Cool intragroup medium and USMg\,II absorbers]{Ultra-strong Mg\,II Absorbers as a Signature 
of Cool Intragroup Gas}
\author[J.-R. Gauthier ]{Jean-Ren\'e Gauthier\thanks{E-mail:jrg@astro.caltech.edu}
\thanks{Based on data gathered with the 6.5 m Magellan Telescopes located at Las Campanas Observatory, Chile and 
at the W.M. Keck Obsevatory, which is operated as a scientific partnership among the California Institute of Technology, 
the University of California, and NASA, and was made possible by the generous financial support of the W.M. Keck Foundation. }\\
Cahill Center for Astronomy and Astrophysics, California Institute of Technology, Pasadena, CA 
}
\begin{document}
\date{}

\voffset=-0.8in

\pagerange{\pageref{firstpage}--\pageref{lastpage}} \pubyear{2012}

\maketitle

\label{firstpage}

\begin{abstract}
We present the results of a spectroscopic survey of galaxies  in the 
vicinity of an ultra-strong Mg\,II $\lambda\lambda 2786,2803$ absorber of rest-frame absorption equivalent 
width $W_r(2796)=4.2$\AA\ at $z=0.5624$. 
This absorber was originally found at projected separation $\rho=246$ kpc of a  luminous red 
galaxy (LRG) at $z=0.5604$. Magellan IMACs spectroscopy has revealed 
two galaxies at $\rho < 60$ kpc ($z=0.5623$ and $z=0.5621$) and a third one at $\rho=209$ kpc ($z=0.5623$) 
near the redshift of the absorber.These findings indicate that the absorbing gas resides in a group environment.
Combining SDSS broadband photometry with additional $B-$, $K_s$-band images and optical 
spectroscopy, we perform a stellar population synthesis analysis of the group members to characterize their 
star formation histories, on-going star formation rates (SFR), and stellar masses. We find that the two group 
members at $\rho < 60$ kpc are  best characterized by old stellar populations ($>1$ Gyr) and little on-going 
star formation activity ($\rm{SFR}<2.9$ \msun/yr), while the third object at $\rho=209$ kpc exhibit 
[O\,II]- and continuum-derived SFR consistent with SFR$>3.0$ \msun/yr. 
Including the two ultra-strong Mg\,II absorbers analyzed by \citet{nestor2011a},  this is the third ultra-strong Mg\,II absorber for which 
a detailed study of the galactic environment is available. All three aborbers are found in galaxy groups.  
We examine different physical mechanisms giving rise to the absorbing gas including starburst driven-outflows, 
cold filaments, extended rotating disks, and stripped gas. We argue that the large equivalent width observed in these absorbers 
is more likely due to the gas dynamics of the intragroup medium rather than driven by starburst outflows.
\end{abstract}


\section{Introduction}

Dense galactic environments constitute a remarkable 
laboratory to study the impacts of dynamical interactions and ram pressure 
stripping on the gaseous content of galaxies and the intragroup medium.  In the local universe, we benefit from 
the sensitivity of 21-cm observations to map cool gaseous structures down 
to column density of H\,I $\rm{N}(\rm{H}\,I) \sim 10^{17}$ \cmjj\ (e.g.,\ \citealt{thilker2004a}).  
These observations revealed that groups host a diversity of H\,I structures, including tails and bridges, 
plumes, disk warps, filaments, and high-velocity 
cloud complexes (e.g.,\ \citealt{sancisi2008a,chynoweth2008a,chynoweth2011a,
mihos2012a,chynoweth2012a,rasmussen2012a}).Among the groups observed locally, the M81/M82 group constitutes a 
remarkable example underlining the complexity of the processes shaping 
up the intragroup medium of galaxy groups. In fact, the M81/M82 radio observations 
presented in \citet{chynoweth2008a} revealed an extended ($\sim 100$ kpc) 
network of gaseous bridges, tails, and isolated clouds with a total 
estimated neutral hydrogen mass of $\approx 10^{10}$ \msun \citep{chynoweth2008a,yun1999a}. 
Similarly extended H\,I features have been found for other groups, 
including M101 \citep{mihos2012a} and NGC 2563 \citep{rasmussen2012a}. Consequently, these structures constitute an 
important reservoir of cool baryons which impact the subsequent evolution 
of the group members. 

The H\,I structures found in groups are thought to arise from a combination 
of several phenomena, including dynamical interactions between group members 
(e.g.\ \citealt{wang1993a}), ram pressure stripping 
(e.g.\ \citealt{kawata2008a,mccarthy2008a}), and possibly in-situ cloud formation 
via thermal instabilities (e.g.,\ \citealt{mo1996a,maller2004a}) . In addition, 
hydrodynamical simulations have shown that accretion of cold gas along 
dark matter filaments (cold mode accretion) is thought 
to be the dominant mechanism of gas accretion for dark matter halos of masses 
$M \lesssim 10^{12}$ \msun , while in more massive systems 
the gas is shock heated to a temperature similar to the virial temperature  
of the dark matter halo. The gas is accreted via ``hot mode" 
accretion. (e.g.,\ \citealt{keres2005a,keres2009a,faucher-giguere2011a,
faucher-giguere2011b}). Although our theoretical understanding of these 
processes has improved significantly over the last decade or so, we
lack empirical data to constrain their relative importance  
in the evolution of groups and galaxies in general.  

While our empirical knowledge of the cool gas content of local groups 
has been primarily shaped by H\,I 21-cm observations, we rely 
extensively on absorption line techniques to trace the extended gas of individual galaxies and groups at $z\gtrsim0.1$. Numerous 
studies have used the Mg\,II $\lambda\lambda$ 2796,2803 absorption 
doublet to trace cool, $T\sim10^4$ K gas with 
$\rm{N}(\rm{H}\,I)\sim 10^{18-22}$ \cmjj\, similar to gas column densities 
detected through 21-cm observations (e.g.,\ \citealt{bergeron1986b,lanzetta1990a,steidel1994a,churchill1996a,churchill2005a,tripp2005a,bouche2007a,chen2008a,kacprzak2008a,menard2009a,barton2009a,chen2010b,gauthier2010a,bordoloi2011a,chen2012a}). 
Although, their physical nature remains debated, several works have 
found that Mg\,II absorbers trace the CGM out to projected distances $\sim 100$ 
kpc from normal, ``isolated" galaxies at $z\lesssim 1.0$ (e.g.\ \citealt{kacprzak2008a,chen2010a,bordoloi2011a}). 

In addition, the cold gas content of massive galaxies and group-size dark matter halos have 
been studied by cross-correlating photometrically-selected luminous red galaxies (LRGs) and Mg\,II absorbers 
found in SDSS at $z\sim 0.5$ \citep{bouche2006a,lundgren2009a,gauthier2009a}. 
Spectroscopic follow-up of close, LRG--Mg\,II pairs by \citet{gauthier2010a} have 
shown that even though the covering fraction, $\kappa$, of $W_r(2796)\geq0.3$\AA\  Mg\,II absorbers 
is much lower around LRGs ($\kappa \approx 0.2$) than normal $\sim L_*$ galaxy ($\kappa \approx 0.7$) at similar redshifts, the 
gaseous envelope extends out to projected distances comparable to the virial radius 
of these massive halos ($\sim 500$ kpc).  The presence/absence of Mg\,II 
absorbers does not correlate with the recent star formation activity of the LRGs, 
strongly suggesting that recent starburst driven outflows from LRGs
are not responsible for the Mg\,II gas \citep{gauthier2011a}. The much larger gaseous envelope around 
LRGs compared to normal, isolated galaxies may also suggest an origin for the 
absorbing gas different from starburst driven outflows . The dark matter halos of LRGs are expected to host approximately five 
satellite galaxies with luminosities $L\gtrsim0.1L_{\rm LRG}$ and within 500 kpc of the LRG \citep{tal2012a,tal2012b}. 
In principle, the Mg\,II absorber could be found in the circumgalactic medium or stripped gas 
of a satellite (e.g.,\ \citealt{wang1993a}), or in correlated, large scale structures seen along the 
line of sight (e.g.\, \citealt{gauthier2010a}). Consequently, a comprehensive analysis of the environment of Mg\,II absorbers 
found in the vicinity of LRGs, including a census of candidate satellite galaxies and a detailed account of their star formation histories 
and on-going star formation rates are necessary to assess the relative importance of these 
physical processes for the production of the observed Mg\,II absorbing gas. 

This paper presents a detailed study of the environment of an ultra-strong, 
$W_r(2796)=4.2$\AA, Mg\,II absorber found at $\rho=246$ kpc and $|\Delta v|=385$ \kms of an LRG at 
$z=0.5604$. Ultra-strong Mg\,II (USMg\,II) absorbers  with $W_r(2796)>3$\AA\ are rare systems harboring 
Mg\,II absorbing gas spread over several 100 km/s\footnote{The minimum velocity width of a saturated 
Mg\,II absorber is $\Delta v_{\rm{min}}=W_r(2796)/$\AA  $\times 107$ \kms}. 
Other similarly strong Mg\,II absorbers have been previously associated with the cold phase of 
starburst driven-outflows based on the symmetry of the absorption profile with respect 
to the galaxy systemic redshift (e.g.,\ \citealt{bond2001a}) or the proximity of the absorbing 
gas to a starburst galaxy \citep{nestor2011a}. 

The USMg\,II absorber studied here is found in a group environment in 
which at least three, normal galaxies with relatively low SFR are found to have redshifts consistent with the 
absorbing gas. Based on the results of a stellar population 
synthesis analysis conducted on the group members, we show that the USMg\,II absorber is unlikely to 
be tracing the cool phase of a recent starburst driven outflow originating from a satellite, but we cannot rule 
out this possibility. Including the two ultra-strong Mg\,II analyzed by 
\citet{nestor2011a},  this is the third ultra-strong Mg\,II absorber for which a detailed study of the 
galactic environment is available. All three aborbers are found in groups of galaxies. We argue that the large 
equivalent width observed in these absorbers is more likely to be due to the gas 
dynamics of the intragroup medium rather than driven by starburst outflows.

This paper is organized as follows. In section 2, we present a series of observations aimed at characterizing 
the environment of the USMg\,II absorber. These observations include $B-$ and 
$K_s$-band images of the QSO field, high-resolution spectroscopy of the Mg\,II absorber, and optical spectroscopic 
follow-up of the group members found in the vicinity of the QSO sightline. In section 3, we discuss the 
methodology and results of a stellar population synthesis (SPS) analysis conducted on the group members. 
The SPS analysis allows us to test the outflow hypothesis by constraining the star formation histories and 
on-going star formation rates of the group members. In section 4, we discuss the implications of our results 
and explore different scenarios for the origin of the absorbing gas.  We adopt a $\Lambda$ cosmology with 
$\Omega_M = 0.3$ and $\Omega_{\Lambda} = 0.7$, and a Hubble parameter $H_0=70~\rm{km/s}/\rm{Mpc}$ 
throughout the paper. All projected distances are in physical units unless otherwise stated. All magnitudes 
are in the AB system. Stellar and halo masses are in units of solar masses. 

\section{Observations and data analysis}

The USMg\,II system presented in this paper was found in the 
spectrum of QSO SDSSJ220702.64$-$090117.8 and was listed 
in the \citet{prochter2006a} catalog. The emission redshift of the QSO 
is $z_{\rm{em}}=1.30$ and the intervening absorber is located at $z_{\rm{Mg\,II}}=0.5624$. 
As part of their comprehensive study to characterize the cool gas content 
of massive dark matter halos, \citet{gauthier2011b} identified a luminous red galaxy (LRG) at  
$\rho=246$ kpc and velocity separation $|\Delta v|=385$ \kms\ from the USMg\,II absorber. 
This projected separation is less than the virial radius ($\approx 500$ kpc) of 
LRGs at $z\sim0.5$ (eg.,\ \citealt{blake2008a,gauthier2009a}) while $|\Delta v|$ 
is comparable to the virial velocity of LRGs suggesting that the 
absorbing gas could be gravitationally bound to the dark matter halo 
of the LRG. 

A first step toward a comprehensive understanding of the physical origin of the 
absorbing gas consists of characterizing the properties of the galaxy populations 
found in the vicinity of the QSO sightline. 
The QSO--LRG field is covered by the SDSS imaging footprint at a depth of 
$i' \approx 22.3$\footnote{Corresponds to a $5-\sigma$ detection of a point source in 
1" seeing at airmass 1.4 \citep{york2000a}.}. The LRG has $i'=19.6$ corresponding 
to $\approx 4 L_*$. Given the typical seeing conditions of $\approx 1.4$", the 
SDSS images become significantly incomplete for detecting faint ($\sim 0.1 L_*$) satellite galaxies 
at $z\approx 0.5$ near the magnitude limit of SDSS. Deeper images are thus necessary 
to reveal the putative satellite population located in the vicinity of the QSO sightline. 

\subsection{$B-$ and $K_s-$band photometry}

We supplemented the already available SDSS $u',g',r',i',z'$ photometry 
with deep $B-$ and $K_s$-band images to identify satellite candidates for 
follow-up spectroscopy.  
We gathered $B-$band images on 2010 September 8 with the 
CCD camera of the du Pont 2.5-m telescope at Las Campanas Observatory. 
The CCD camera consists of a single 2048$\times$2048 detector of pixel 
size 0.259" with a field-of-view of 8.8' sqr arcminute. The observations were carried 
out in a series of 3 exposures of 600 s each and were spatially dithered by an amount 
varying from 10 to 50 arcsec. The images were gathered under photometric 
conditions. The FWHM of the point spread function (PSF) in the final combined image is 
$\approx$ 1.2". We used the $g'$-band photometry and the $(g'-r')$ color of 
$\approx 10$ stars the field to compute the 
$B-$band photometric zero point (e.g.,\ \citealt{chonis2008a}).

Data reduction followed standard procedures. We subtracted the bias of 
individual frames by using the overscan region of the chip. Next, the 
bias-subtracted frames were flattened using a combined dome flat. 
We then registered individual exposure using bright stars found 
in the field, filtered for cosmic rays, and combined the images. 
The combined images reach a $3-\sigma$ magnitude limit of 25.6 in a circular 
aperture of radius 1". Photometry and source extraction were performed using SExtractor v2.5 
\citep{bertin1996a}. The image was convolved with a 1.3" FWHM 
gaussian kernel for object detection. We also required a continuous area 
of 5 pixels and a minimum threshold of 3$\sigma$ above noise level. 

In addition, $K_s$-band images were obtained with the near-IR imaging spectrograph MOSFIRE \citep{mclean2010a}
on the Keck I telescope. In imaging mode, the field of view of MOSFIRE is 6.1'$\times$6.1' 
with a plate scale of 0.18"/pix. 
The images were obtained on the night of 18 October 2012 by C. Steidel. 
The observations consist of a series of 10 exposures. Each exposure is a sum of ten 
8.7s co-adds yielding a total of 87s per exposure. The total on-target exposure time is 870s. 
Each co-adds were dithered by  30" to 40". The data were taken under 
photometric conditions. 

The data were reduced by Allison Strom using the IRAF XDIMSUM package and custom 
IDL routines. The data were flat-fielded using the background sky signal from a combination 
of the science exposures. Images were registered using point sources and shifted to a common 
reference system. The sky background was subtracted using the IRAF task XSLM which is part 
of the XDIMSUM package. Thirteen unsaturated 2MASS point sources were recovered over the entire frame 
yielding a zero point of 28.49 mag. The PSF in the final combined 
image is characterized by FWHM$\approx$ 0.63". The combined image reaches a $3\sigma$ magnitude limit 
of 23.6 over a circular aperture of radius 1". SExtractor parameters were the same as for the 
$B-$band dataset, except that we convolved the image with a 0.6" FWHM gaussian kernel 
for source detection. 
For both $B-$ and $K_s$-band observations, the magnitudes quoted in this paper correspond to 
SExtractor elliptical-aperture measurement MAG\_AUTO. 

In Figure 1, we show the $K_s$-band image of the field. The left panel displays the 
positions of the QSO and the LRG. The inset on the right shows 
a zoomed-in view of the vicinity of the QSO. We subtracted the QSO PSF in the inset panel. 
We labeled galaxies with letters (A to J) and, when available, we included the redshift derived 
from IMACs spectroscopy (see \S 2.2). We put yellow boxes around the galaxies with 
redshifts consistent  with the absorber ($z_{\rm Mg\,II}=0.5624$). Galaxies  A,B,C,D, and J 
are detected at $\rho < 60$ kpc from the QSO sightline. We were able to obtain reliable redshift estimates for 
galaxy A and B.  These galaxies are found at velocity separation $|\Delta v|<80$ \kms\ from the absorber 
and are members of a group including galaxy G at $\rho=209$ kpc. Since galaxies 
A,B, and G are found within the projected virial radius and virial velocity of the LRG, we consider 
the LRG as a likely member of the group. Nonetheless, it is possible that the LRG and the group formed by galaxies A, B, and G 
are inhabiting distinct dark matter halos, implying that their relatively small separations arise from mere projection effects. 
It is also possible that the LRG is accreting a subgroup composed of galaxies A,B, and G as seen in nearby poor 
groups (e.g.,\ \citealt{zabludoff1998b}). When estimating the halo mass 
of the group (see \S 3.1), we considered both scenarios in which the LRG is a member or inhabits a different 
halo. We found that excluding the LRG from the group membership yields similar halo mass and does not affect 
the conclusions of this paper. Hereafter, we will consider the LRG to be a member of the group unless otherwise stated. 
In columns (1)-(6) of Table 1, we list the galaxy ID, spectroscopic redshift, 
SDSS photometric redshift, $B-$ and $K_s$-band photometry and estimated luminosities of the confirmed 
group members as well as galaxies C, D, and J. In column (10) and (11) we included the projected separation $\rho$
from the QSO sightline and the velocity separation $\Delta v$ from the redshift of the Mg\,II absorber measured 
in SDSS ($z_{\rm Mg\,II}=0.5624$) . 
 
 \begin{figure*}
\centerline{
\includegraphics[angle=0,scale=0.35]{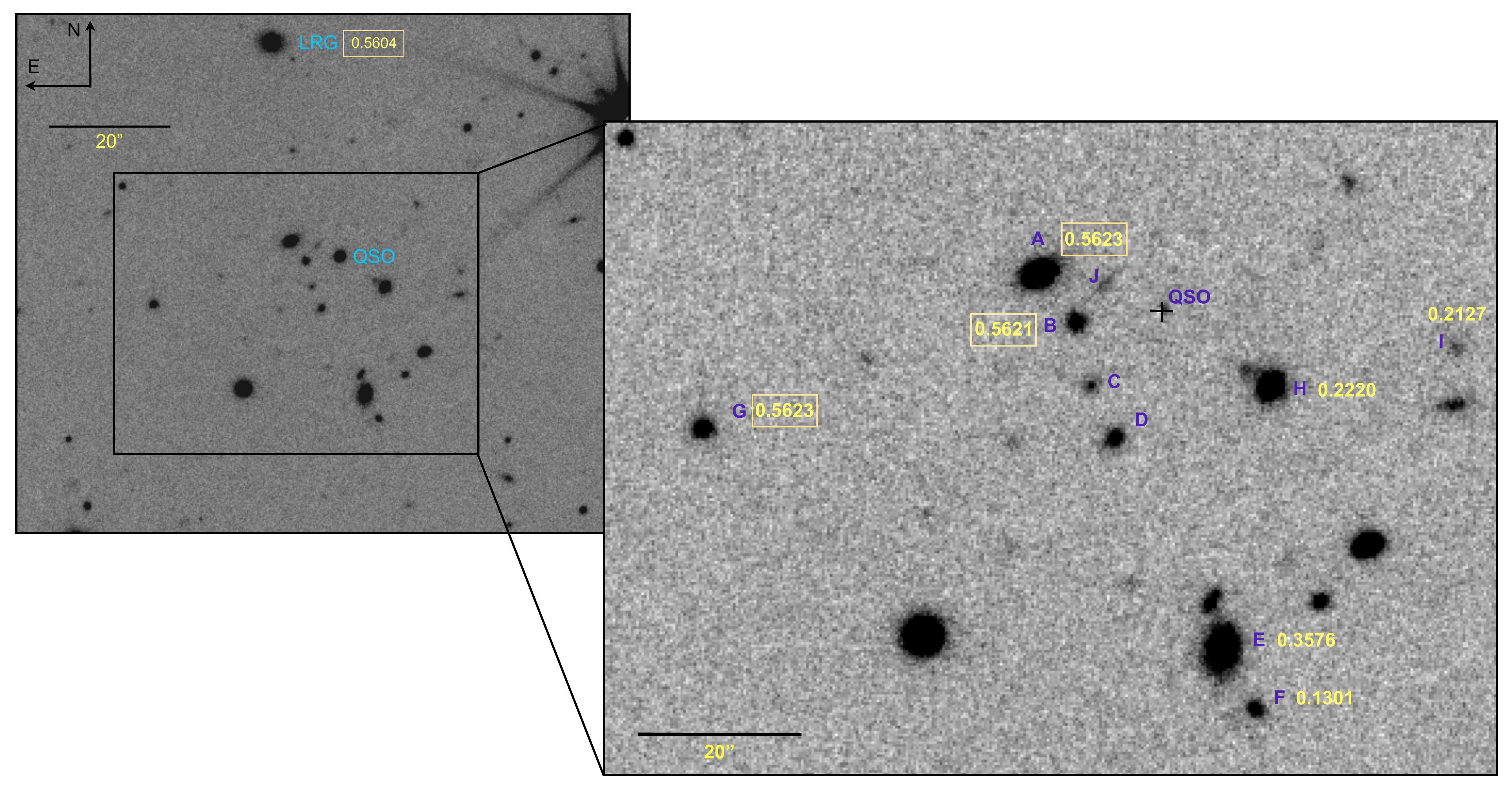}
}
\caption{$K_s-$band image of the LRG--QSO field obtained with MOSFIRE on the Keck I telescope. In the left image, we highlight the position 
of the LRG with respect to the QSO sightline. The LRG is located at $\rho=246$ kpc from the QSO.  In the inset image on the right 
we show the vicinity of the QSO sightline after subtracting the QSO PSF. The location of the QSO is marked by the black cross. We obtained 
optical spectra of galaxies A to I  and reliable redshifts were found for galaxies  A, B, E, F, G, H, and I.  Galaxies A ($1.8 L_{{K}^*}$), B ($0.3 L_{{K}^*}$), and G 
($0.5 L_{{K}^*}$) have redshifts consistent with the USMg\,II ($z_{\rm Mg\,II}=0.5624$) and are denoted by yellow boxes.}
\label{model}
\end{figure*}
 
\subsection{Follow-up spectroscopy with IMACs}
Follow-up optical spectroscopy of the galaxies
was carried out on the night of 19 August 2012 with the IMACs imaging 
spectrograph \citep{dressler2011a} and the f/2 camera. We conducted the 
observations with a long slit of width 1.2". Given the serendipitous alignment 
of the galaxies, we were able to obtain spectra of galaxies A--I with only 
two long-slit orientations. Galaxies A--F were observed with the first long-slit 
configuration, while the second long-slit orientation allowed us to 
simultaneously obtain spectra of galaxies G--I. For all spectroscopic 
observations, we used the 200 l mm$^{-1}$ grism which offers a spectral coverage of 
$\lambda = 5000-9000$\AA\ with $\approx$ 2\AA\ per-pixel resolution. The 
observations were carried in a series of 4$\times$1800s exposures 
for the first slit configuration and a single 1200s exposure for the second orientation. Observations 
of He-Ne-Ar lamps were performed after each series of two science exposures 
for accurate wavelength calibrations. The typical seeing was $\approx 0.8$". 

The spectra were processed and reduced using the Carnegie 
Observatories COSMOS program\footnote{Available at http://code.obs.carnegiescience.edu/cosmos}.
COSMOS is based on a precise optical model of the spectrograph which allows 
for an accurate prediction of the locations of the slit and spectral features. Initial guesses are 
further refined after using known spectral lines obtained through He-Ne-Ar lamps observations. 
The science frames were bias-subtracted and flat-fielded following standard procedures. 
Sky subtraction on individual 2-D spectrum was performed following the \citet{kelson2003a} 
procedure. Optimal weights based on the variance of each pixel were used to extract the 
1-D spectra. The spectra were then calibrated to vacuum wavelengths, corrected 
for the heliocentric motion and for Galactic extinction using the \citet{schlegel1998a} maps. 
Finally, the 1-D spectra and associated error arrays were flux-calibrated 
using a sensitivity function derived from observations of standard star EG21. 
Redshifts of the galaxies were determined based on a cross-correlation analysis 
using known SDSS templates. 

In Figure 2, we display the spectra of all galaxies with enough $S/N$ to yield a reliable redshift 
estimate. We divided the figure into group  
members and foreground interlopers. Because our flux calibration is uncertain at $\lambda \gtrsim 8700$\AA, 
we limited our analysis to $\lambda \leq 8700$\AA. 
The LRG spectrum was obtained with the B\&C spectrograph on the du Pont telescope 
as described in \citet{gauthier2010a}. 
Overlaid on top of each spectrum are SDSS $r'$ and $i'$ 
photometric datapoints in red. Galaxy A and the LRG exhibit spectra features dominated 
by absorption transitions, indicating an old underlying stellar population and little star formation 
in the recent past. Galaxies B and G show nebular emission lines of [O\,II] and H$\beta$, while [O\,III] is 
present in galaxy G indicating recent star formation activity. Unfortunately, parts of the H$\beta$ 
emission line of galaxies B and G fall in the A-band absorption of atmospheric O$_2$.  
Galaxies B and G also exhibit a shallower SED slope indicating younger stellar populations. 
In columns (8) and (9) of Table 1, we list [O\,II] luminosities and a rough estimate of the star formation 
rate based on the \citet{kennicutt1998b} [OII] calibrator. No slit-loss or intrinsic dust extinction corrections were applied to $L([\rm{O\,II}])$. 

\begin{figure*}
\centerline{
\includegraphics[angle=0,scale=0.65]{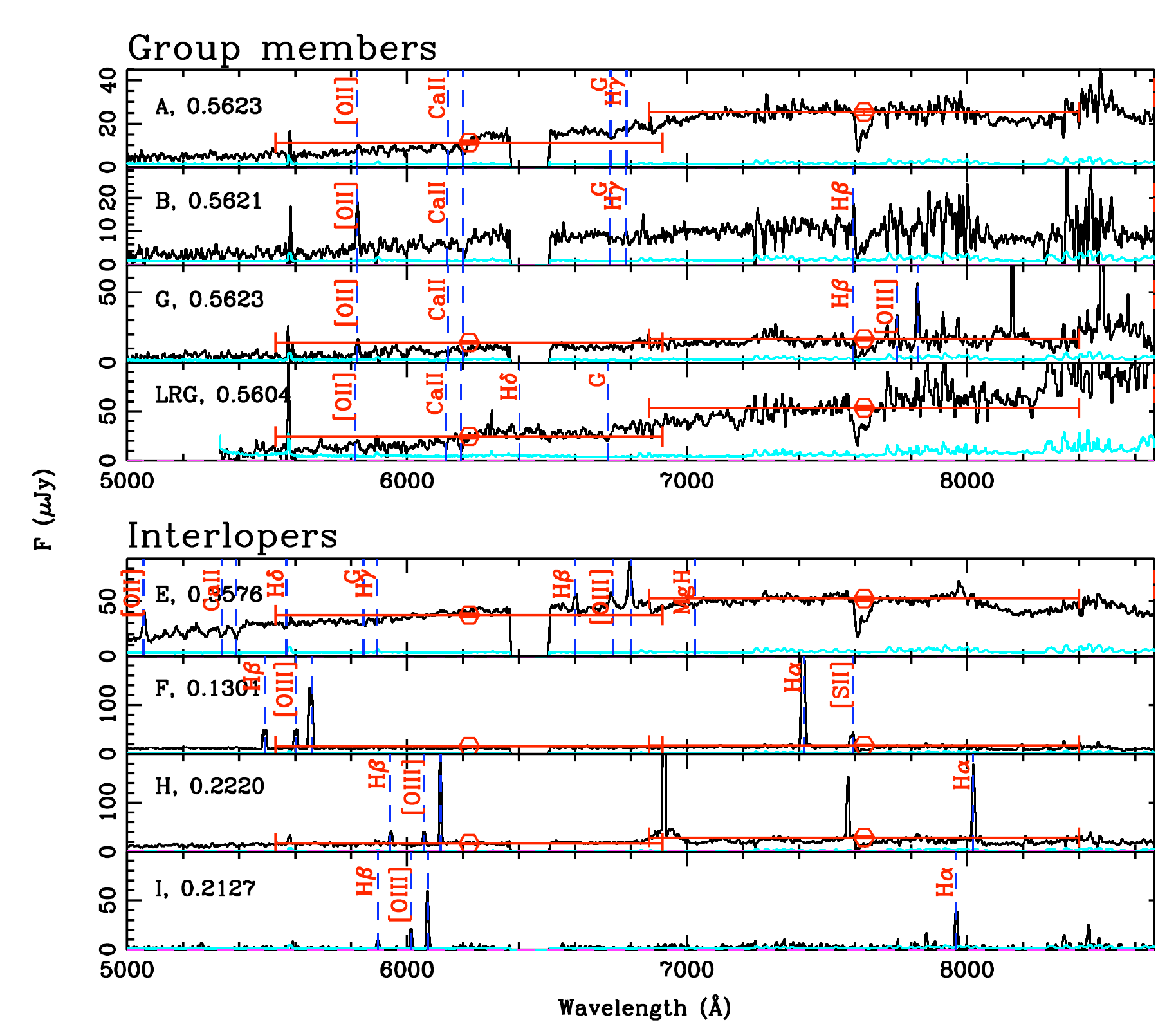}
}
\caption{Optical spectra of the group members along with the interloper systems detected in the 
vicinity of the USMg\,II absorber at $z=0.5624$. We show the spectra in black along with the error array in cyan. 
All spectra were obtained with IMACs on Magellan except for the LRG whose spectra was acquired with the 
B\&C spectrograph on the du Pont telescope (see \citealt{gauthier2010a} for 
more details). We labeled each plot with the object name along with its spectroscopic redshift. 
The open red points and error bars correspond to the SDSS $r'$ and $i'$ photometry. The red error bars correspond to the 
FWHM of the filter passbands. Note that the spectra were smoothed with a 3-pixel wide top-hat kernel for display purposes. }
\label{model}
\end{figure*}
 
\subsection{Echelle spectroscopy of the USMg\,II absorber}
At the resolution of SDSS spectra ($\approx 150$ \kms), it is impossible to resolve the velocity 
field of individual absorbing clumps which giving rise to the USMg\,II 
absorber. 
Echelle spectroscopy is thus necessary to obtain the velocities 
of the absorbing clumps relative to the group members and compare, for example, 
the resulting kinematics with predictions from supergalactic wind models.

We obtained high-resolution spectra of the QSO on the night of 6 September 2010
with the MIKE spectrograph \citep{bernstein2003a} 
located on the East Nasmyth platform of the Magellan Clay telescope. MIKE offers wavelength coverage 
$\lambda=$3200-9500 \AA\ by using a dichroic that redirects 
the light into a blue ($\approx$ 3200-4800 \AA\ ) and red ($\approx$ 
4400-9500 \AA\ ) channel.  We used MIKE standard configuration, a 2$\times$2 binning mode, and a 1.0" slit.  
The blue channel has a dispersion of $\approx$ 0.04 \AA\ $/\rm{pix}$ in a 2$\times$2 binning mode.
The 1" slit provides a measured resolution of $\approx$ 10.7 \kms\ in the blue channel where the Mg\,II absorber 
resides.   The observations consist of $3\times3000$s exposures 
and no dither was applied between exposures. Flat-field sky flats for the 
blue channel were obtained at twilight. Calibration frames (Th-Ar) 
for wavelength solutions were taken immediately after each science exposure using the internal 
calibration lamps. The MIKE observations are characterized by a typical seeing of 0.9" and were 
obtained under mostly photometric conditions. The data were reduced using the IDL package MIKE 
Redux\footnote{Available at http://web.mit.edu/~burles/www/MIKE/} written by S. Burles et al.  The spectra 
were calibrated to vacuum wavelengths and corrected for heliocentric motion. The final 1-D spectra was binned 
using 3 \kms wide pixels and has $S/N$ (per pixel) $\approx$ 6 in the spectral region where the 
Mg\,II absorber is located. The continuum level was determined by fitting a series of 3rd-order b-splines in the spectral 
regions of interest. 
\begin{figure}
\centerline{
\includegraphics[angle=0,scale=0.50]{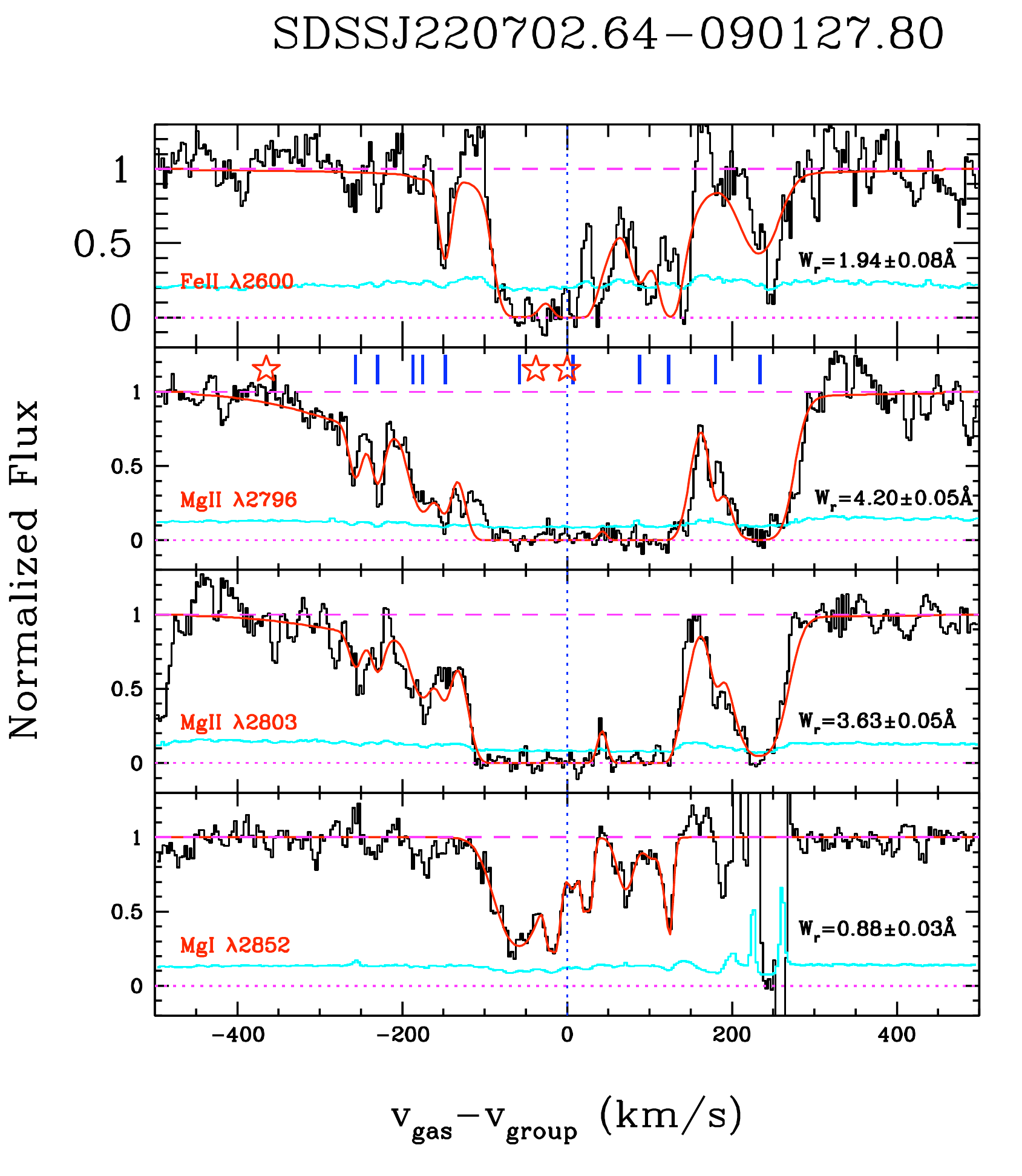}
}
\caption{Line-of-sight velocity distribution of absorbing clumps. The spectrum has been smoothed with a 3-pixel wide top hat kernel 
for display purposes. In each panel, the absorption spectrum is shown in black histogram 
with the 1-$\sigma$ error in cyan. In addition to Mg\,II, we also observe strong Fe\,II $\lambda 2600$ and Mg\,I $\lambda 2852$ 
transitions. We included the rest-frame equivalent width measurements for each transition. Note that the error on $W_r$ does not 
include the uncertainty on the continuum level. A Voigt profile analysis of the observed Fe\,II, Mg\,II, and Mg\,I absorption profile 
yields a minimum of 11 individual absorption components. The best-fit models are shown in red curves and the position of 
individual components are also marked by blue tick marks at the top of the Mg\,II $\lambda2796$ transition panel. Zero velocity 
in each panel corresponds to the group members A,  B, and G light-weighted redshift ($z_{\rm{group}}=0.5623$). We indicated the 
positions of galaxies A, B, G, and the LRG with red stars in the top of the Mg\,II $\lambda 2796$ transition panel. The absorbing 
clumps display relative line-of-sight motions ranging from $\Delta v=-257$ to $\Delta v=+234$ \kms\ with respect to the group systemic redshift. 
 }
\label{model}
\end{figure}

We searched for all absorption transitions within the blue channel of MIKE that are associated with 
the USMg\,II absorber. These include Fe\,II $\lambda 2600,2586,2382,2374$, Mn\,II $\lambda 2576,2594$, 
Mg\,I $\lambda2852$, and Fe\,I $\lambda 2484$ transitions. 
Unfortunately, the QSO is faint ($r'=18.6$) for echelle spectroscopy and Fe\,II transitions blueward 
of Fe\,II $\lambda 2600$ have too low $S/N$ to yield reliable constraints while the Mn\,II transitions were 
simply not detected. We thus limited our analysis to Mg\,II, Mg\,I, and Fe\,II $\lambda2600$ transitions. 
We performed a Voigt profile analysis that considers all the observed absorption features at once, 
using the VPFIT\footnote{Available at http://www.ast.cam.ac.uk/~rfc/vpfit.html} software 
package. We considered the minimum number of components required 
to deliver the best $\chi^2$ in the Voigt profile analysis. Finally, we established the 
observed line-of-sight velocity by comparing the relative velocities of individual component 
with the $K_s-$band light-weighted redshift of group members A, B, and G ($z_{\rm{group}}=0.5623$).  
In this case, we adopted a conservative approach and excluded the LRG from the redshift estimate of 
the group. A smoothed  version of the absorption profile is show in Figure 3 along 
with the best-vpfit model shown in red ($\chi^2_r \approx 2$). In the Mg\,II $\lambda2796$ panel, 
we display the positions of individual components with blue tick marks and the positions of galaxies A, B, G and 
the LRG with red stars. The gas has line-of-sight velocities ranging from $\Delta v= -257$ to $+234$ \kms\  
with respect to $z_{\rm{group}}$.

\section{Stellar population synthesis analysis}
 To directly test the previously found 
association between USMg\,II absorbers and the cold phase of recent starburst-driven outflows 
(e.g.,\ \citealt{nestor2011a}), we performed a stellar population synthesis analysis to constrain 
the star formation histories, stellar masses, and on-going star formation rates of group members  
A, B, G, and the LRG.

To accomplish this task, we carried out a likelihood analysis that compares the optical spectrum 
and broad-band photometry of the galaxies with model expectations for different 
stellar age ($t$), metallicity ($Z$), star formation history ($\tau$), and intrinsic dust extinction ($\tau_V,\mu$). The likelihood 
function follows
\begin{equation}
\mathcal{L}(t,Z,\tau,\tau_V,\mu) = \prod_{i=1}^N \exp \Bigg \{ -\frac{1}{2} w_i \Bigg[ \frac{f_i - \bar{f}_i(t,Z,\tau,\tau_V,\mu)}{\sigma_i}   \Bigg]^2  \Bigg \} 
 \end{equation}
where $N$ is the number of
spectral bins ($N=1823$ for galaxies A,G, $N=1820$ for B, and $N=1228$ for the LRG), 
$f_i$ is the observed flux in the $i$th bin,  $\bar{f}_i$ is the model prediction, $\sigma_i$ is the
corresponding error of the $i$th element, and $w_i$ is the statistical weight associated with 
each element. The weights $w_i$ corresponds to the width of each spectral 
element in wavelength space.
In the calculation of the likelihood function, we consider the 
broadband photometric data points as spectral elements. The weight of each photometric 
datapoint  corresponds to the FWHM of the filter passband. In the range $\lambda \approx 5000-8700$\AA, 
the photometric data points $g',r',i',z'$ overlap in wavelength space with the optical spectra. Since the same 
part of the SED is probed by both broad-band photometry and spectroscopy, we divided the 
weights of each overlapping spectral element by two. We adopted this procedure because 
in the spectral regions where both spectroscopy and broadband photometry overlap, broadband 
photometry does not provide additional constraints on the SED. Consequently, we avoid giving more 
weight to parts of the SEDs that were probed twice. 

The stellar population models were based on those described in \citet{bruzual2003a} revised 
to include a prescription of the TP-AGB evolution of low and intermediate mass stars \citep{marigo2007a}. 
We employed a Chabrier initial mass function for all models \citep{chabrier2003a}.  The star 
formation history (SFH) of the model galaxies was parametrized by either a single burst (ssp) or by 
an exponentially declining model with an $e$-folding timescale $\tau$. The ages $t$ were equally 
separated in logarithmic space between $10^5$ and 8 Gyr, where the upper-limit corresponds 
to the age of the Universe at $z=0.56$. We also adopted an equal spacing of 50 Myr for $\tau$ from 
0.1 to 0.5 Gyr and we adopted metallicities of 0.005, 0.02, 0.2, 0.4, 1, and 2.5 solar. To directly 
compare between data and models, we convolved the model spectra 
with a top-hat function of width 350 \kms to mimic the resolution of the spectra. We also applied 
the extinction curve of \citet{charlot2000a} to simulate dust attenuation by the host galaxies. 
The extinction model is characterized by $\tau_V$, the total effective $V-$band optical depth 
affecting stars younger than 10$^7$ yr and $\mu$, the fraction of $\tau_V$ contributed by the 
diffuse ISM. For each model, we generated a grid of extinction curves. Each extinction curve was 
characterized by a value of $\tau_V$ and $\mu$. We adopted $\tau_V=0,0.2,0.5,1.0,1.5$, and 2.0 
and $\mu=0,0.25,0.5,0.75$ and 1.0.  The range of $\tau_V$ is consistent 
with galaxies ranging from dust-free early-types to obscured star-forming systems \citep{dacunha2008a}. 
In total, 1380 models were generated. 

In the left panels of Figure 4 we show the spectra of the LRG and galaxies A,B, and G in black, 
along with the SDSS $u', g', r', i', z'$ and $B-$,$K_s-$band photometric data points in blue. The 
best-fit stellar population model is shown in magenta. In the upper-left corner, we labelled the name 
of each galaxy along with its redshift and the best-fit model parameters.  The LRG and galaxy 
A exhibit a very red SED  with $B-K_s>4$ with very little or no nebular emission lines (see Figure 2), suggesting an old 
underlying stellar population and little star formation in the recent past. 
Galaxies B and G have bluer $B-K_s$ color, indicating younger stellar populations and star formation 
activity. In the case of A and the LRG, the best-fit models are old 
($t=1.3$ and 2.6 Gyr respectively) and characterized by solar and sub-solar metallicities 
(1.0 and 0.4 $Z_{\odot}$) . The best-fit $e-$folding times are $\tau=0$ (ssp) and 0.5 Gyr respectively. 
In contrast, the best-fit models of galaxies B and G reveal younger stellar populations with 
$t=0.6$ and 0.05 Gyr respectively, short $\tau$ (0.3 and 0.2 Gyr) and sub-solar metallicities. 

The presence of younger stellar populations in galaxies B and G is further supported 
by the results of the likelihood analysis presented in the right panel of Figure 4. In this panel, we 
show the relative likelihood of a given stellar age to reproduce the galaxy SED. To compute 
the relative likelihood, we identified the model with the largest $\mathcal{L}$ 
at a given age and computed its relative likelihood defined as $\exp( \mathcal{L}_{\rm max}-\mathcal{L})$, 
where $\mathcal{L}_{\rm max}$ is the maximum value of the likelihood function. The right panel of Figure 4 
shows that models as young as 10 Myr are able to reproduce the SED of galaxies B and G, 
although these models also require large dust extinction ($\tau_V=2$). This is not the 
case for galaxy A and the LRG where models younger than $\approx$ 1 Gyr are rejected. 
At $t\lesssim$ 300 Myr, the models reproducing B and G typically have $\tau \lesssim 0.2$ Gyr, 
solar or sub-solar metallicity and large dust extinction ($\tau_V=2$ and $\mu=1$). Older models typically have longer 
$\tau$ ($\tau \sim 0.2-0.5$ Gyr), lower metallicities ($\lesssim$ Z$_{\odot}$) and lower dust extinction 
($\tau_V=0-0.5,\mu=0.5-1$).  At $t \lesssim 3$ Gyr, the LRG and galaxy A are best modeled by single 
burst and $\tau \lesssim$ 0.3 Gyr, super-solar metallicities, and large dust extinction ($\tau_V \geq 1$). 
Alternatively, older models generally have larger $\tau$ and solar metallicity. 

\begin{figure*}
\centerline{
\includegraphics[angle=0,scale=0.6]{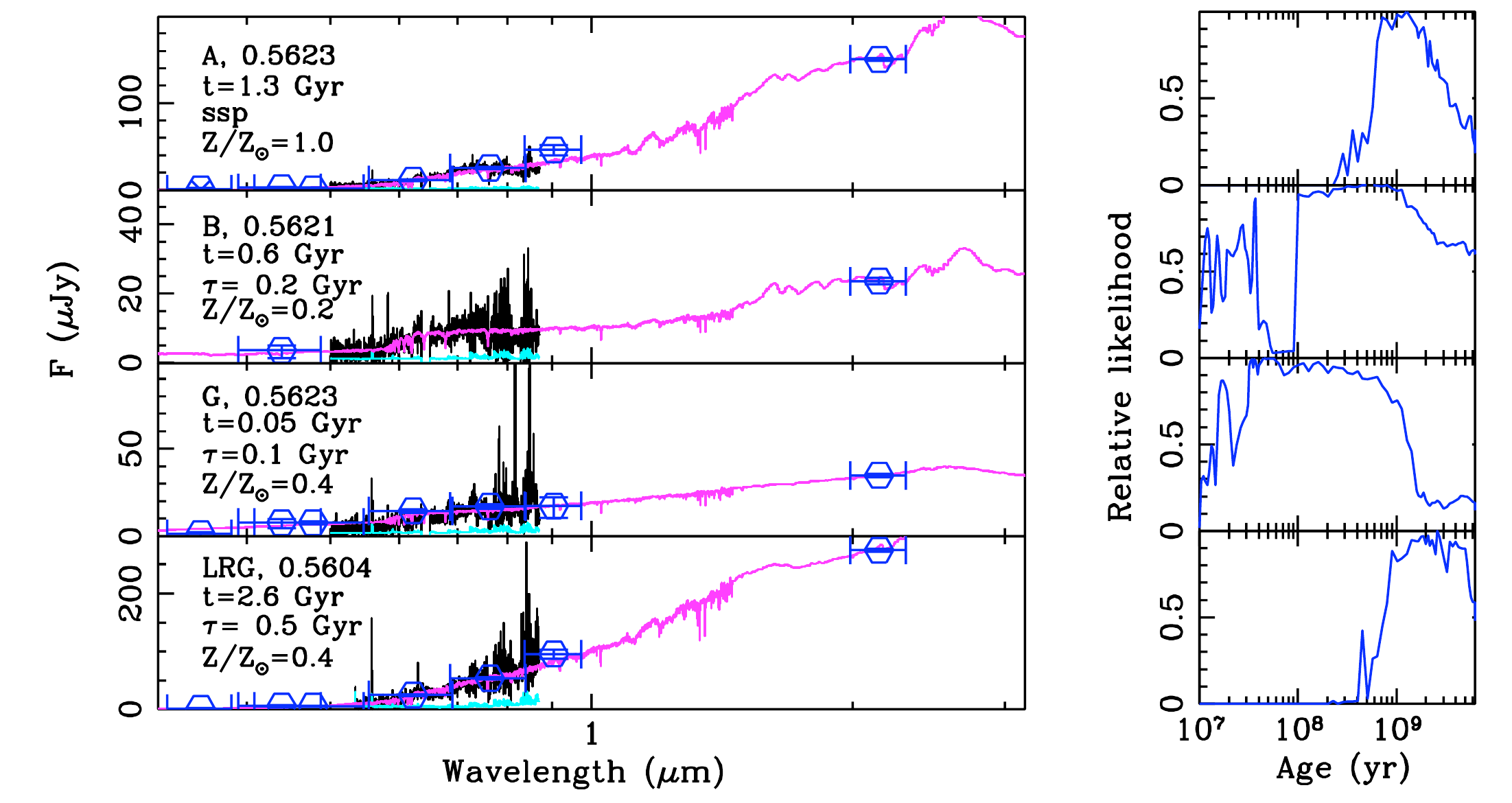}    
}
\caption{Stellar population synthesis analysis of the group members A,B,G, and the LRG. In the left panel, 
we show the spectra of each galaxy in black, along with the error array in cyan and the broadband photometric 
datapoints in blue. For all objects, $B-$ and $K_s$-band datapoints are shown and for galaxies A, G and the LRG, 
SDSS $u', g', r', i'$ and $z'$ are also displayed. The best-fit SED model is shown by the magenta line. We wrote the 
best-fit model parameters in each panel. The right panel shows the relative likelihood functions of the stellar age 
of the galaxies. Galaxy A and the LRG are characterized by old ($>1$ Gyr) stellar populations with little on-going 
star formation. Galaxies B and G are characterized by younger and dustier models.
}
\label{model}
\end{figure*}

 \subsection{Estimated galaxy properties}
   
Galaxies B and G have younger stellar populations and 
could, in principle, host starburst-driven outflows, although their [O\,II]-derived SFR listed 
in Table 1 indicate low levels of recent star formation activity. To obtain a continuum-derived, 
SFR estimate for galaxies B and G, we first selected all models with relative likelihood $>$0.5.
While this threshold is arbitrary, the selected 
models are representative of the variations in star formation history, metallicity, and intrinsic 
dust extinction that are allowed by the data. For each selected model and galaxy, we 
computed the stellar mass using rest-frame $K_s$-band photometric data point and the 
mass-to-light ratios $M/L_{K_{s}}$ derived 
from the model. The continuum-derived SFR (SFR$_{\rm cont}$) was computed by multiplying the SFR model prediction 
by the ratio of the galaxy stellar mass over the model stellar mass. 

The stellar mass and SFR$_{\rm cont}$ distributions of galaxies A, B, G, and the LRG 
are shown in Figure 5. The number of models with a relative likelihood $>$ 0.5
varies from $N_m=204$ for galaxy A to $N_m=2339$ for galaxy B. Because of the degeneracy between stellar age and 
dust attenuation, a much wider range of models are able to reproduce the relatively flat SEDs of galaxies B and G than the steep SED slope 
of galaxy A. We found that galaxies B and G have $\log \rm{M}_{*} \sim 9.5$ which is similar to the stellar masses 
of $L_*$ galaxies at $z\approx0.5$. In contrast, galaxy A and the LRG are more massive with $\log \rm{M}_*=10.6^{+0.1}_{-0.2}$ and 
$10.9^{+0.3}_{-0.2}$ respectively. Using the abundance matching technique of \citet{moster2010a} and assuming 
that galaxy A is the central object, we found\footnote{We converted the stellar 
mass for a Kroupa IMF which is the IMF used in \citet{moster2010a}. $M_{\rm{Kroupa}}=1.12M_{\rm{Chabrier}}$.} 
that the dark matter halo hosting the group has $\log \rm{M}_h = 12.3^{+0.1}_{-0.2}$. If one assumes that the LRG is the 
central object, the halo mass increases by 0.4 dex to $\log \rm{M}_h = 12.7^{+0.6}_{-0.3}$, a value consistent with 
the dynamical mass estimate of  groups at $z\lesssim 1.0$ (e.g.,\ \citealt{balogh2011a}). We included  the stellar masses of each 
galaxy along with their 1-$\sigma$ error bars in column (7) of Table 1.

In the right panel of Figure 5, we show the distribution of the SFR$_{\rm cont}$ for all four galaxies. 
For galaxy G, this distribution is characterized by a tail extending out to SFR$_{\rm cont} \sim 100$  \msun/yr. 
For this reason, we quoted the lower limit of the 84\% C.I..  
The distribution of $\rm{SFR}_{\rm cont}$  confirms the
old nature of the stellar populations of galaxies A and the LRG. These galaxies, including galaxy B 
have $\rm{SFR}_{\rm cont} <2.9 $\msun/yr. A more interesting case is galaxy G with  SFR$_{\rm cont}>3.0$ \msun/yr (84\% C.I.).   
The median value of galaxy G's distribution is $\rm{SFR}_{\rm cont} = 31$ \msun/yr. This is 
an order of magnitude larger than $\rm{SFR}_{\rm [O\,II]}$. Several factors could explain the lower $\rm{SFR}_{\rm [O\,II]}$ value, 
including slit loss and dust attenuation. In addition, large $\rm{SFR}_{\rm cont}$ values occur in model with $\tau_V \geq 1.5$ 
implying large dust attenuation corrections $>10$. In column (12) of Table 1, we listed the limits to the  84\% 
confidence interval of SFR$_{\rm cont}$ for all four galaxies. 

Since we do not have high enough quality spectra
and SDSS photometry to constrain the SEDs of galaxies C, D, J, 
we derived their stellar masses assuming that they are the redshift of the USMg\,II absorber 
and that the combined distribution of mass-to-light ratios $M/L_{K_s}$ derived for 
galaxies A,B,G, and the LRG is an adequate proxy for galaxies C, D, and J. 
In essence, we hypothesized that these galaxies share similar 
star formation histories with the confirmed group members.
The stellar masses of galaxies C, D, and J are listed in column (7) of Table 1. The results of this 
analysis show that if located at the redshift of the absorber, galaxies C,D and J are sub-$L_*$ systems
with $\log \rm{M}_* \sim 9$. 

\section{Discussion}
We performed a comprehensive analysis of the environment of a $W_r(2796)=4.2$\AA\ 
Mg\,II absorber at $z=0.5624$. Deep optical and near-IR images have 
revealed five absorber host candidates within 60 kpc of the QSO sightline. Follow-up 
optical spectroscopy showed that two galaxies at $\rho < 60$ kpc and one 
at $\rho=209$ kpc have redshifts consistent with the absorber. The three confirmed group 
members have rest-frame $K_s$-band luminosities ranging $\approx 0.3-1.8$ $L_{K^*}$. 
These observations convincingly demonstrate that the Mg\,II absorber originates in a 
group environment consisting of at least three galaxies. Whether or not the LRG is 
part of this group remains an open question. We adopted a conservative approach 
and included the LRG in the analysis, but we found that the conclusions of this paper 
are not affected by this decision.  

We conducted a stellar population synthesis analysis on the group members to constrain 
their stellar masses, star formation histories, and SFR. 
In addition to the SFR inferred from SED models, we also derived a SFR estimate 
based on [O\,II] emission. Both techniques give consistent results for galaxies A, B, and the LRG. 
We found that the star formation rates of galaxies A, B, and the LRG are low ($<2.9$ \msun/yr). 
Furthermore, the SEDs of galaxy A and the LRG are best characterized by old ($>1$ Gyr) stellar 
populations confirming the passive nature of these systems. Of the three group members, only galaxy G exhibits 
significant star formation rate with a continuum-derived value of SFR$_{\rm cont}>$ 3 \msun/yr.  This value 
is consistent with the [O\,II]-derived one (SFR$_{\rm [O\,II]}=2.5\pm1.1$ \msun/yr). Note that we applied 
no slit loss or dust extinction corrections when estimating SFR$_{\rm [O\,II]}$. 
 
\subsection{The supergalactic wind scenario} 
 
According to \citet{heckman2001a}, galaxies with star formation rate 
per unit area exceeding $\Sigma_{\rm SFR} = 0.1$ \msun/yr/kpc$^2$ show ubiquitous starburst-driven outflows. 
We estimated $\Sigma_{\rm SFR}$ for the group members by adopting the $K_s-$image isophotal area as a 
proxy for the sizes of the galaxies. All galaxies are spatially resolved in the $K_s$-band images.
For galaxies A and B located at $\rho < 60$ kpc, we found $\Sigma_{\rm SFR} < 4\times10^{-3} $ 
\msun/yr/kpc$^2$ and $\Sigma_{\rm SFR} < 0.02$ \msun/yr/kpc$^2$ respectively. These estimates are based on the 
upper limits of the SFR$_{\rm cont}$ distribution 84\% confidence interval. For the LRG, $\Sigma_{\rm SFR}$ is even 
smaller, $\Sigma_{\rm SFR} < 5\times10^{-4}$ \msun/yr/kpc$^2$. Therefore, it is unlikely that 
galaxies A, B, and the LRG are launching galactic superwinds at the time of observation. 
In contrast, galaxy G has SFR$_{\rm cont}>3$ \msun/yr. Adopting the median value of the SFR$_{\rm cont}$ 
distribution ($\widetilde{\rm{SFR}}_{\rm cont}=31$ \msun/yr) as a typical value allowed by the SED models, we found 
$\Sigma_{\rm SFR} \approx 0.13$ \msun/yr/kpc$^2$. This value is above the minimum 
threshold for launching superwinds. In principle, the USMg\,II absorber could be tracing 
the cold phase of a starburst-driven outflow originating from galaxy G. Recent semi-analytical models 
of galactic superwinds have shown that a combination of radiation and ram pressures could drive cold 
entrained gas out to distances $\gtrsim 100$ kpc \citep{murray2011a}. The distance traveled by the 
cold clumps depend on the physical properties of the galaxy as well as the mass load parameter ($\beta$)
and the fraction of supernova luminosity that is thermalized ($\epsilon$). Under the starburst-driven wind scenario, 
the equal distribution of Mg\,II absorbing gas on both sides of the systemic redshift of galaxy G would imply 
that G is either very inclined or that the opening angle of the wind material is very large \citep{gauthier2012a}. 

In contrast,  \citet{chen2010a} analyzed a sample of $\approx 100$ QSO-- isolated galaxy pairs at $z\sim 0.25$, 
and convincingly demonstrated an anti-correlation between $W_r(2796)$ and $\rho$. According to their 
results,  USMg\,II with $W_r>3$\AA\ are expected to be found within $\approx 11$kpc of a typical 
$L_*$ galaxy. According to their results, the USMg\,II absorber is more likely to be associated with galaxies A and B 
at $\rho < 60$ kpc than galaxy G at $\rho=209$ kpc. For a definite assessment of the wind scenario we are planning 
to obtain obtain $HST$ images of the field, allowing us to use the galaxy inclination and QSO--galaxy 
relative orientation to estimate the de-projected kinematics of the Mg\,II clumps and 
constrain the acceleration mechanisms of the putative wind \citep{gauthier2012a}. 
 
 \begin{figure}
\centerline{
\includegraphics[angle=0,scale=0.9]{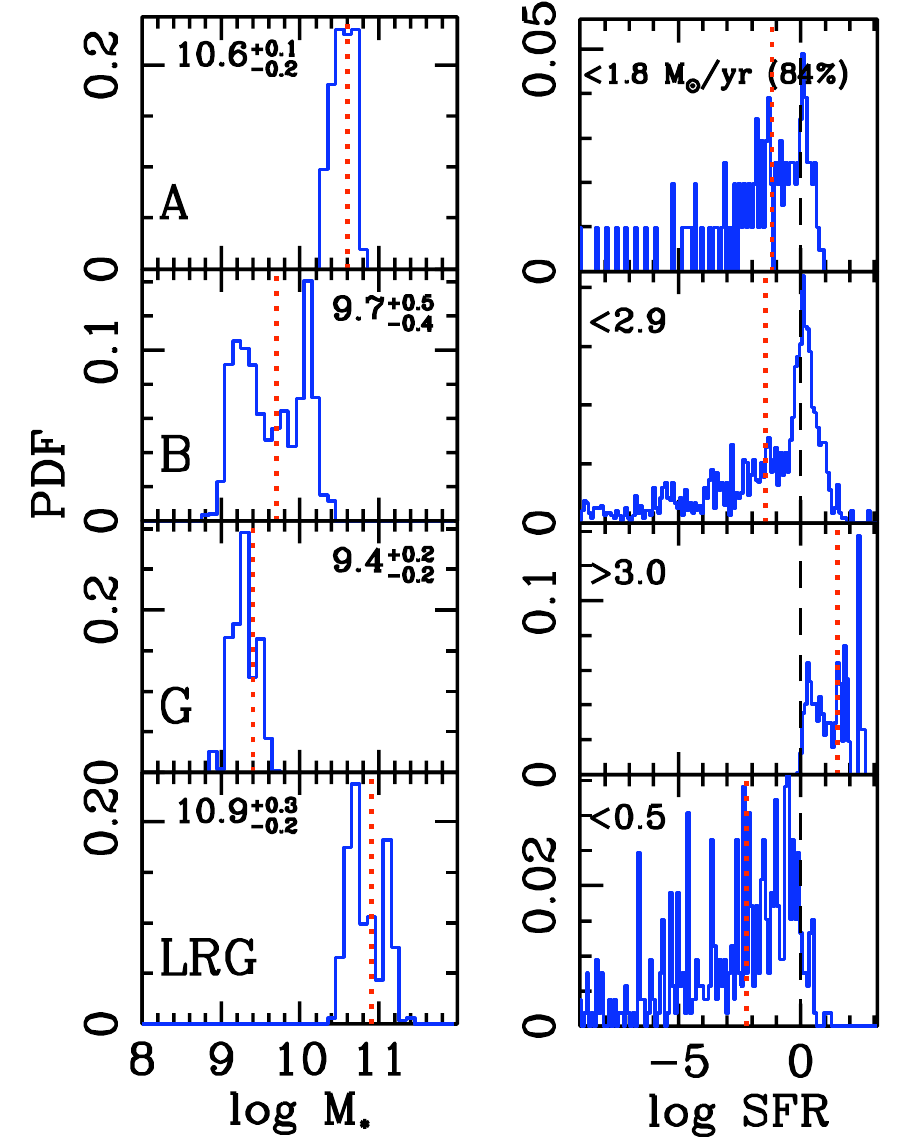}   
}
\caption{\emph{Left:} stellar mass distribution inferred from the SED models with relative likelihood $>$0.5. In each panel, 
we labelled the median value along with the error bars corresponding to the 16-84\% confidence intervals (C.I.) 
of the distribution. Galaxies B and G have stellar masses similar to $L_*$ galaxies at $z\sim0.5$ while galaxy A and the LRG are much more massive. The dotted red line denotes the 
median value. \emph{Right:} Continuum-derived log SFR (SFR$_{\rm cont}$) distribution inferred from the same models. 
For galaxies A, B, and the LRG, we labelled the value corresponding to the upper-limit defining the 84\% C.I. of the SFR$_{\rm cont}$ distribution. For galaxy G, we labelled the lower-limit of the 84\% C.I.. The median value of the distribution ($\widetilde{\rm SFR}_{\rm cont}$) is also represented by a vertical dotted red line. We also added a dashed vertical line at 1 M$_{\odot}$/yr to guide 
the eye.  
For galaxies A, B, and the LRG, the median values are $\widetilde{\rm SFR}_{\rm cont}<0.1$ \msun/yr. In contrast, galaxy G has $\widetilde{\rm SFR}_{\rm cont}=31$ \msun/yr. Such large SFR$_{\rm cont}$ may be enough to drive starburst driven outflows (see \S 4). In the case 
of galaxy G, the models producing SFR$_{\rm cont}>$10 \msun/yr are young ($<10^{8}$ yr) and very dusty ($\tau_V\geq1.5$).
}
\label{model}
\end{figure}

\subsection{The \citet{nestor2011a} sample of USMg\,II absorbers }
 
 In a recent paper, \citet{nestor2011a} presented deep $r-$band observations, follow-up spectroscopy, 
 and stellar population synthesis analysis of the host candidates of two USMg\,II at $z\approx 0.7$. In 
 both fields, the authors found two $\gtrsim L_*$ galaxies within 60 kpc of the QSO sightline and with 
 redshifts similar to the absorber. These results also indicate that a  group environment is associated with 
 both absorbers. Assuming that the most massive galaxies are the central objects and following 
 \citet{moster2010a}, we computed the halo masses of these two USMg\,II hosts based 
 on the estimated stellar masses published in \citet{nestor2011a}.  In Table 2, we list 
 the stellar and derived halo masses these two USMg\,II along with the system presented in this paper. 
 In columns (6)-(7) we also list the number of confirmed group members their respective separations 
 from the QSO sightline.  All three Mg\,II absorbers are found in halos with $\log M_h > 12$. 
 Their estimate of the attenuation-corrected SFR showed that 
 the galaxies display significant star formation activity (SFR$ \gtrsim 9$ \msun/yr). 
 Although \citet{nestor2011a} acknowledged the possibility that the Mg\,II gas 
 originate from dynamical interactions, they attributed the USMg\,II systems to the 
 cold phase of a starburst-driven outflow. 
 
 These conclusions are further supported by the star formation rate surface density 
 of the galaxies associated with the two USMg\,II systems. 
 We computed $\Sigma_{\rm SFR}$ for all galaxies associated with the two USMg\,II systems presented 
 in \citet{nestor2011a}. We adopted the largest dust-corrected values as the SFR of the galaxies while the size of each 
 galaxy was derived from the sizes of the isophotal ellipses presented in \citet{nestor2007a}. We did not attempt 
 to de-project the sizes of the galaxy disks. Instead these values allow us to assign upper 
 limits to $\Sigma_{\rm SFR}$. Note that \citet{nestor2007a} obtained $i'-$band images for these two fields while 
 our galaxy size estimates are based on $K_s$-band images. We found that one galaxy in each field has $\Sigma_{\rm SFR}$ 
 that could be above the threshold of 0.1 M$_{\odot}$/yr/kpc$^{2}$. These two galaxies are G07-1 ($\Sigma_{\rm SFR} < 0.17$) and 
 G14-1 ($\Sigma_{\rm SFR} < 0.45$). \citet{nestor2011a} argue that the  
 small probability of finding galaxies with such extreme SFR in the vicinity of a rare USMg\,II system implies that 
 the two are related. However, we do not find such star-forming systems at $\rho < 60$ kpc of the USMg\,II 
 absorber discussed in this paper. Yet all three systems are found in group environments. 
 
 In addition to the sample of USMg\,II absorbers presented in \citet{nestor2011a}, a detailed study of the 
 environment of a strong, $W_r(2796)=1.8$\AA\ Mg\,II absorber was presented in \citet{kacprzak2010b}. 
 The authors found that the absorber inhabit a group composed of at least five sub-$L_*$ galaxies showing moderate 
 star formation activities of a few M$_{\odot}$/yr.  Furthermore, \citet{whiting2006a} conducted a spectroscopic survey of 
 galaxies found in the field toward QSO PKS2126-158. They discovered a group of at least five passive galaxies at the 
 redshift of a $W_r(2796) \approx 2.5$\AA\ Mg\,II absorber at $z=0.66$. These studies show that finding galaxy 
 groups associated with strong Mg\,II absorbers may not be an uncommon occurrence. 
 
 \subsection{Different scenarios for the origin of the gas}
 
 In principle, the QSO sightline could intercept the CGM of multiple, correlated 
structures along the line of sight giving rise to the USMg\,II absorber (e.g.,\ \citealt{pettini1983a}). 
According to this hypothesis, the combined effect of multiple gaseous halos would produce 
absorbing gas spread over several 100 \kms\ as observed in 
USMg\,II absorbers. As mentioned before, the anti-correlation between 
$W_r(2796)$ and $\rho$  implies that 
$W_r(2796)>1$\AA\ are typically found at $\rho \lesssim 30$ kpc of isolated $L_*$ galaxies 
(e.g.,\ \citealt{chen2010a,steidel1994a}). For a 0.1$L_*$ system similar to galaxy J 
located at $\rho=28$ kpc\footnote{If the galaxy has $z=z_{\rm Mg\,II}$.}, strong Mg\,II absorbers 
with $W_r(2796)>1$\AA\ would be limited to even smaller impact parameters ($\rho \lesssim 13$ kpc). None of galaxies C, D, and J 
meet this criterion. 

Other works have directly compared the kinematics of strong, $W_r\gtrsim 1.0$\AA\ absorbers with  
the kinematics of galaxies found within $\sim 100$ kpc of the QSO sightline 
(e.g.,\ \citealt{steidel2002a,kacprzak2010a,bouche2012b}). In most cases, the absorbing gas lies entirely 
on one side of the galaxy systemic redshift and exhibits velocities consistent with the galaxy's rotation curve, suggesting 
that a significant fraction of the absorbing gas reside in an extended rotating disk 
\citep{steidel2002a,kacprzak2010a}. This scenario alone cannot explain the gas kinematics we measured 
for the USMg\,II absorber. In fact, the gas is found on both sides of the systemic redshift of all confirmed 
group members. These observations suggest that other physical mechanisms are needed to explain 
the velocity separation between the gas and galaxies and the overall velocity spread of the gas. 
Because the quality of our images forbids a reliable characterization of the galaxy morphology, we cannot rule out the 
possibility that the gas is found in an extended rotating disk associated with galaxy A,B,C, D, or J. However
the observed line-of-sight velocity spread ($\approx 500$ \kms) would imply a very massive disk which 
is highly improbable for $\lesssim L_*$ galaxies. 

Recent hydrodynamical simulations of galaxy formation have shown that cold gas can reach the 
central regions of dark matter halos via a filamentary accretion channel (e.g.,\ \citealt{keres2009a,stewart2011a}). 
In these dense filaments, the gas is never shock-heated and can penetrate deep in the hot coronal 
envelope of the dark matter halo. However, the covering fraction of optically thick gas 
originating in the cold filaments is found to be low ($<$5\% -- see \citealt{fumagalli2011a})  
and the efficiency of the cold-mode accretion 
depends strongly on halo mass and redshift. For halos with $\log M_h \gtrsim 12$, most 
of the gas accretion occurs in a hot mode where the gas temperature is similar to the virial temperature 
of the halo. Furthermore, the metallicity of the cold filaments is likely to be low, rendering the detection 
of metal line transitions challenging (e.g.,\ \citealt{kimm2011a}). For these reasons, it is rather unlikely 
that the USMg\,II absorber is found in the cold streams of gas seen in simulations.  

Alternatively, we argue that the Mg\,II absorber resides in the stripped gas once bound to the 
gravitational potential of the group members. As discussed in the introduction, such 
stripped material is seen in nearby, gas rich,  group environments. Furthermore, if the USMg\,II absorber 
is tracing stripped gas, the kinematics of the absorbing clumps should reflect, 
the range of velocities allowed by the gravitational potential of the dark matter halo of the group. As show in 
Figure 3, absorbing clumps have $|\Delta v|\leq 257$ \kms\, which is comparable to the central velocity 
dispersion ($\sigma_v \approx 180$ \kms) of a $\log M_h=12.5$ NFW dark matter halo 
at $z=0.5$. The lack of current star formation activity in galaxies A and B may also indicate that  cold gas 
has been stripped from these galaxies . Further \emph{HST} images of this field would provide 
crucial morphological informations on the group members and would allow us to directly test the 
stripped gas hypothesis for the origin of the USMg\,II absorber population. 

\section*{Acknowledgments}

It is a pleasure to thank H.-W. Chen, M. Rauch, J. Mulchaey, 
T. Tal, K. Cooksey, A. Diamond-Stanic, G. Becker and  C. Steidel
for helpful comments and discussions.  We also thank 
the referee for helpful comments that improve the draft significantly. 
JRG gratefully acknowledges the financial support of a Millikan 
Fellowship provided by Caltech. This work was supported by the National 
Science Foundation through grant AST-1108815. This paper is 
dedicated to the memory of Caltech Professor W.L.W. Sargent.

\onecolumn
\footnotesize{
\begin{landscape}
\begin{table*}
 \centering
 \begin{minipage}{140mm}
  \caption{Galaxy Properties.}
  \begin{tabular}{@{}cccccccccccr@{}}
  \hline
   ID & redshift & photo-$z$\footnote{SDSS ``photozcc2" value. } & $B$ & $K_s$ & $L_K/L_{K*}$\footnote{ $M_{K^*} - 5\log h = -22.07$ (AB) from the UKIDSS UDS galaxy sample  of Cirasuolo et al. (2010) .} & $\log$ M$_{*}$\footnote{Stellar mass obtained from SED models with relative likelihood$>$0.5. For galaxies C,D, and J, we assumed that they are located at the redshift of the absorber and adopted the likely models of galaxies A,B, and G to estimate their stellar masses.}  &  $L([\rm O\,II])$ & SFR$_{[\rm{OII}]}$\footnote{Star formation rate derived from $L([\rm{O\,II}])$ using the \citet{kennicutt1998b} estimator. No slit losses or intrinsic dust attenuation corrections were applied.} & $\rho$ & $\Delta v$ & SFR$_{\rm{cont}}$\footnote{For galaxies A,B, and the LRG, the value corresponds to the upper-limit defining the 84\% C.I. of the SFR derived from SED models. For galaxy G, we quoted the lower-limit of the 84\% C.I.. }  \\
   & & & & & & & ($\times 10^{40}$ ergs/s) & ($\rm{M}_{\odot}$/yr) & (kpc) & (km/s) & ($\rm{M}_{\odot}$/yr)  \\ 
(1) & (2) & (3) & (4) & (5) & (6) & (7) & (8) & (9) & (10) & (11) & (12) \\   
   \hline
   A &  0.5623 &  & 23.0$\pm$0.6  & 18.46 $\pm$ 0.01 & 1.8 & $10.6^{+0.1}_{-0.2}$   & $3.6\pm0.4$& $0.5\pm0.2$ & 55 & -38 &  $<1.8$  \\ 
    B &  0.5621 &    & 22.7$\pm$0.5 & 20.47 $\pm$ 0.04 & 0.3 & $9.7^{+0.5}_{-0.4}$   & $18\pm1$& $2.5\pm0.8$ & 38 &  -77 & $<2.9$ \\ 
   C & & & 23.4$\pm$0.6 & 21.81 $\pm$ 0.10 &  0.09 & $8.9^{+0.7}_{-0.3}$ & & & 45 & &  \\
   D &  &  & 23.8$\pm$0.9 & 20.54 $\pm$ 0.05 & 0.3 & $9.4^{+0.7}_{-0.3}$ &  &  & 59  &  \\    
   G & 0.5623 & 0.31 $\pm$ 0.12 & 21.7$\pm$0.4 & 20.05 $\pm$ 0.03  & 0.5 & $9.4\pm0.2$ & $18\pm2$& $2.5\pm1.1$ & 209 & -38 & $>3.0$   \\ 
   J & & & $<$25.6 & 22.00 $\pm$ 0.14 & 0.07 & $8.8^{+0.7}_{-0.3}$ & & & 28 & \\
   LRG & 0.5604 & 0.53 $\pm$ 0.04 & 22.0$\pm$0.4 & 17.80 $\pm$ 0.01 & 3.5 & $10.9^{+0.3}_{-0.2}$ & $<9$ (3$\sigma$) &$<1.3$ & 246 & -385  &  $<0.5$ \\  
 \hline
\hline
\end{tabular}
\end{minipage}
\end{table*}
\end{landscape}
}
\twocolumn

\begin{table*}
 \centering
 \begin{minipage}{140mm}
  \caption{The environment of USMg\,II absorbers at $z\lesssim1$}
  \begin{tabular}{@{}lccccccc@{}}
  \hline
  Reference & $z_{\rm Mg\,II}$ & $W_r(2796)$ & $\log$ M$_*$\footnote{Stellar mass of the most massive group member.} & $\log$ M$_h$\footnote{Halo mass obtained from \citet{moster2010a} assuming that the most massive galaxy is central.} &  $N$ & $ \rho $    \\
   & & (\AA) & & & & (kpc)  \\
   \hline
   \citet{nestor2011a} &  0.7646 & 3.63$\pm0.06$ & $11.23^{+0.32}_{-0.12}$ & $13.3^{+0.6}_{-0.3}$  & 2 & [36,61]     \\
   \citet{nestor2011a} &  0.6690 & 5.6$\pm0.5$ & $10.34^{+0.24}_{-0.16}$ & $12.0^{+0.3}_{-0.1}$  & 2 & [29,58]    \\
   This work & 0.5624 & 4.20$\pm0.05$ & 10.6$^{+0.1}_{-0.2}$ & 12.3$^{+0.1}_{-0.2}$ & 3 & [38,55,209]   \\
 \hline
\hline
\end{tabular}
\end{minipage}
\end{table*} 

\end{document}